\documentclass[prl,twocolumn,super,groupedaddress,superscriptaddress]{revtex4}
\usepackage{graphicx}%
\usepackage{dcolumn}
\usepackage{amsmath}
\usepackage{latexsym}
\usepackage{amssymb}
\begin{document}
\title{Effect of interactions on the cellular uptake of nanoparticles}
\author{Abhishek Chaudhuri}
\email{a.chaudhuri1@physics.ox.ac.uk}
\affiliation{Department of Biomedical Science, University of Sheffield, Western Bank, Sheffield S10 2TN, United Kingdom}
\affiliation{The Rudolf Peierls Centre for Theoretical Physics, University of Oxford, 1 Keble Road, Oxford OX1 3NP, United Kingdom}
\author{Giuseppe Battaglia}
\affiliation{Department of Biomedical Science, University of Sheffield, Western Bank, Sheffield S10 2TN, United Kingdom}
\author{Ramin Golestanian}
\affiliation{The Rudolf Peierls Centre for Theoretical Physics, University of Oxford, 1 Keble Road, Oxford OX1 3NP, United Kingdom}

\date{\today}
\begin{abstract}
We present a simple two-state model to understand the size-dependent
endocytosis of nanoparticles.
Using this model, we elucidate the relevant energy terms required to
understand the size-dependent uptake mechanism and verify it by correctly
predicting the behavior at large and small particle sizes. In the absence
of interactions between the nanoparticles we observe an asymmetric distribution
of sizes with maximum uptake at intermediate sizes and a minimum size
cut-off below which there can be no endocytosis.
Including the effect of interactions in our model has remarkable effects on
the uptake characteristics. Attractive interactions shift the minimum size
cut-off and increase the optimal uptake while repulsive interactions make
the distribution more symmetric lowering the optimal uptake.
\end{abstract}


\maketitle

\section{Introduction}
\noindent
The endocytic process \cite{mayor,doherty} is of paramount importance to
understanding the cellular
uptake of nano-materials, essential for the development of gene and
targeted drug delivery tools. A key feature in the development of such tools
is to achieve effective cytosolic delivery.
To this purpose there have been experiments using liposomes \cite{lasic},
nanoparticles (NPs) \cite{hu}, polymerosomes \cite{beppe1,beppe2,beppe3},
nanotubes \cite{kostarelos,strano1},
electroporation \cite{neumann} and ultrasonic
treatments \cite{kim}. Several of these techniques also suffer
from the problem of high levels of cytotoxicity although recent experiments
using polymerosomes overcome this shortcoming.

In several of these experiments using gold and silver
nanoparticles, nanotubes and polymerosomes \cite{aoyama1,aoyama2,aoyama3,
chitrani1,chitrani2,jiang,beppe1,beppe2,beppe3,strano1} particle size
plays an important role in the cellular uptake. These experiments suggest
that endocytosis of NPs is receptor-mediated and that there is an optimal
size where the uptake is maximum. Most theoretical approaches
\cite{tzlil,suresh,gao,bao} to study the effect of NP geometry on
cellular uptake predict a threshold radius
below which there can be no cellular uptake, and an asymmetric distribution
of the uptake which decays with particle size. Although these approaches
correctly predict the size where the uptake is optimal ($\sim 20-30$ nm),
experimentally the distribution is symmetric and does not seem to
agree with the lower bound as predicted from the theories \cite{chitrani1,chitrani2}.
The answer to this anomaly could be hidden in the highly complex endocytic
mechanism itself.

The endocytic process involves the selection and segregation of the cargo at the cell
surface, subsequent invagination and pinching off from the cell membrane,
and, finally, the transport of these vesicles to intracellular compartments where
they fuse with the target membrane. The mechanisms by which specific cargo
are internalized differ in their morphological and
biochemical details \cite{mayor,doherty,marsh}.
However, recent evidence \cite{johannes} suggests the need to consider
the sharing of molecular machinery depending on the nature of the cargo and to
understand the basic physical principles common to these different uptake
mechanisms.

A key step in the endocytic process is the segregation and clustering
of cargo on the cell membrane which are believed to be the sites where
molecular machinery may be recruited to generate membrane curvature,
form membrane invaginations and subsequently cause scission
\cite{mayor,johannes}. The mechanisms of
formation of these nanodomains can be both passive and active. Although
passive clustering which does not involve ATP hydrolysis can occur via
intermolecular cargo interactions, Reynwar et. al. \cite{reynwar}
showed using coarse grained simulations, that curvature-inducing model
proteins adsorbed on lipid bilayer membranes could experience attractive
interactions that occur purely as a result of membrane curvature.
These interactions could result in clustering and subsequent invaginations.
Thus passive clustering can happen even in the absence of specific cargo
interactions. An example of the passive clustering is the binding of
Shiga toxin---a bacterial toxin---to glycolipid receptors, Gb3,
in the cell membrane of certain cell types \cite{ewers,romer}.
Although the Shiga toxin molecules do not interact directly,
they induce the clustering of the Gb3 lipids, thereby causing local
membrane curvature.

The active mechanisms for cell surface clustering requires ATP and are
therefore energy dependent processes. An example of active clustering is the
formation of nanoscale clusters of glycosylphosphatidylinositol anchored proteins
(GPI-AP) on the cell membrane which are
required for their subsequent endocytosis \cite{sharma}.
These proteins exist as monomers
and nanoclusters ($\sim 4-6$ nm in size, consisting of $< 5$ molecules) with
the interconversion between the two being spatially heterogeneous, being
coupled to an active cortical cytoskeleton. Perturbing the cortical actin
activity affects the construction, dynamics and spatial organization of these
nanoclusters \cite{goswami}.
This type of active segregation also happens with ganglisides
GM1 and GM3 in the exoplasmic (outer) leaflet \cite{fujita} and
Ras isoforms in the cytoplasmic (inner) leaflet of the plasma membrane
\cite{plowman}.

The above examples of passive and active clustering do not involve the
clathrin mediated endocytic pathway where the entrapment of the cargo
occurs by its association with adapter proteins. However, even in the clathrin
coat mediated endocytic pathway, the clathrin lattice organizes the epsins
or BAR domain proteins into domains, which then locally deform the
membrane \cite{ford,henne}.
Therefore the segregation and clustering of cargo on the cell surface is
highly important in the endocytic process. This naturally raises
the following questions: could cell surface clustering affect
the size-dependent cellular uptake of NPs and if so how could we model the
clustering process? Recent experiments \cite{strano1} have shown evidence
of NP surface clustering on the cell membrane and it is important to investigate
this in some detail.

Here, we study systematically for the first time, the effect of
interactions on the cellular uptake of NPs using a thermodynamic model first
proposed by Tzlil et al. \cite{roux,tzlil} and subsequently studied by
Zhang et al. \cite{suresh} in the NP context. We develop our
model by incorporating interactions between NPs. Using a simplified
two-state version of the model, we then elucidate the relevant energy terms
which affect the uptake of NPs in the absence of interactions. We then show
that interactions between NPs indeed affect the minimum radius of uptake
as well as the distribution.

\begin{figure*}[t]
\begin{center}
\includegraphics[height=6cm,width=15cm]{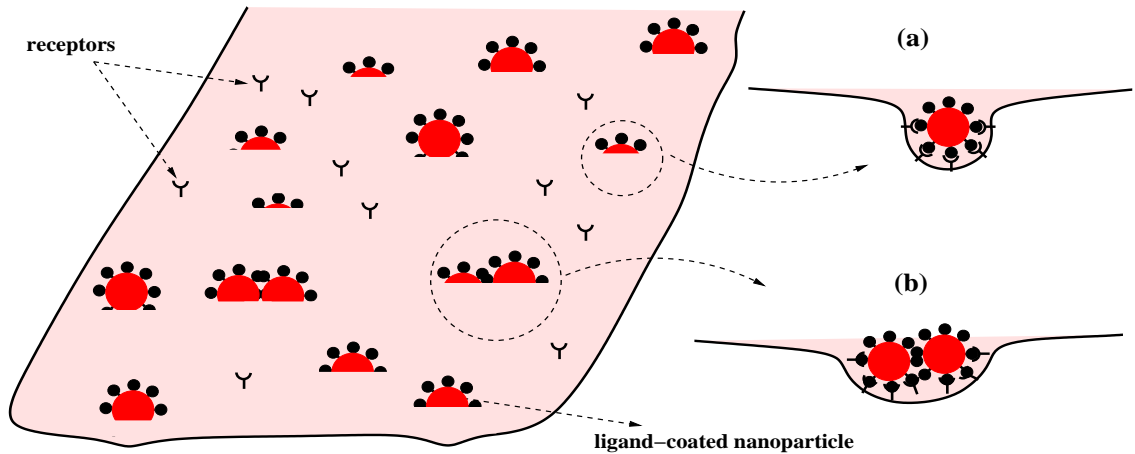}
\end{center}
\caption{Schematic figure of adhering NPs wrapped by cell membrane. The
NPs are wrapped to different degrees with some of them being internalized.
(a) Single NP wrapping showing ligand-receptor binding. (b) Cluster of NPs
wrapped by the cell membrane.}
\label{size-dist}
\end{figure*}

\section{The Model}
\noindent
In this model, the system, which consists of a cell and ligand-coated spherical
NPs in a solution, is in a thermodynamic equilibrium at which a certain
number of NPs are endocytosed. At this state, $N$ NPs adhere to the cell
surface that contains $L$ receptors via ligand-receptor binding, and are wrapped
to different extents by the cell membrane (see Fig. \ref{size-dist}). The
receptors diffuse freely on the cell surface and are segregated into $L_p$
free receptors in the planar membrane and $L_b$ bound receptors in the
curved regions. Let $A$ be the cross-sectional area of a receptor.
For a given NP radius $R$, the number of receptors that can
attach to the NP is $K = 4\pi R^2/A$. For convenience, we shall choose $A$
as our unit of area and $\sqrt{A} = R\sqrt{4\pi/K}$ as our unit of length.
The total membrane area is denoted by $MA$ (in units of
$A$) and is therefore the total number of sites on the membrane that are
accessible to the receptors. The surface concentration of NPs is then 
$c = N/MA$.

We have assumed that the time scale for endocytosis is much larger than the
time for the receptors to diffuse and segregate into the curved and planar
regions. Therefore the distribution of wrapping sizes and receptor densities
can be treated using equilibrium statistical mechanics. 

Let $n_k$ denote the number of NPs wrapped by a membrane section of
area $k$ where $k$ varies discretely between $k=0$ (unwrapped state) and
$k=K$ (completely wrapped state). Then, we have
\begin{equation}
N = \sum_{k=0}^{K} n_{k},\label{eq:N}
\end{equation}
and
\begin{equation}
M_{b} = \sum_{k=0}^{K} k n_{k},
\end{equation}
where $M_bA$ is the total curved membrane area associated with the wrapped
NPs and $M_pA = (M - M_b)A$ is the total area of the planar regions.
The binding of a ligand and a receptor releases chemical energy, $\epsilon$,
which drives the wrapping at the cost of the energy required to bend the membrane.
Therefore, the diffusion of receptors inside the curved regions should lower
the energy of the system facilitating wrapping. However, this leads to the
segregation of receptors between planar and curved regions, which costs
entropy. Also diffusion of free receptors into the curved regions
increases the total curved area (more ligand-receptor bonds) and hence
increases the total membrane bending energy. Furthermore, attractive
(repulsive) interactions between the NPs could lead to clustering
(anti-clustering) and therefore affect the wrapping size distribution of
NPs. To determine the size distribution of the varyingly wrapped NPs, we first
write down the free energy of the system as,
\begin{eqnarray}
\nonumber
\frac{\cal{F}}{k_B T} &=& M_{p}\left[\phi_p\ln{\phi_p}+ (1 - \phi_p)\ln(1 - \phi_p)\right]
\\
\nonumber
&+& M_{b}\left[\phi_b\ln{\phi_b} + (1 - \phi_b)\ln(1 - \phi_b)\right] \\
\nonumber
&+& \sum_{k} n_{k}[\ln \left(n_k/M\right) - 1] - \epsilon L_b + \hat{\kappa}M_b \\
&+& \sum_{k} n_k\Lambda_k + \sum_{k} n_k\Gamma_k + w\sum_{k,k^{\prime}}kk^{\prime}n_kn_k{\prime},
\label{free}
\end{eqnarray}
where $\phi_p = L_p/M_pA$ and $\phi_b = L_b/M_bA = (L - L_p)/M_bA$ denote the
densities of the receptors in the planar membrane and the wrapped regions
respectively, $k_B$ denoting the Boltzmann constant and $T$ being the temperature.

The first three terms in the free energy are entropic contributions written in
terms of a two dimensional lattice gas model:
\begin{itemize}
\item $M_{p}[\phi_p\ln{\phi_p} + (1 - \phi_p)\ln(1 - \phi_p)]$ represents the
configurational entropy of $L_p$ free receptors distributed in the $M_p$
sites of the planar parts of the membrane.

\item $M_{b}[\phi_b\ln{\phi_b} + (1 - \phi_b)\ln(1 - \phi_b)]$ represents the
configurational entropy of distributing $L_b$
receptors among the $M_b$ sites of the curved regions.

\item $\sum_{k} n_{k}[\ln \left(n_k/M\right) - 1]$ is the configurational entropy of a 2D mixture
of wrapped NPs when treated as a multicomponent ideal gas. However, we are
interested in interacting NPs and would therefore have to include interaction
energy for this 2D mixture.
\end{itemize}

The next five terms are energetic:

\begin{itemize}
\item $-\epsilon L_b = -M_b\phi_b\epsilon = -\phi_b\epsilon\sum_{k} kn_k$ is the
total chemical energy released upon the binding of $L_b$ ligand-receptor pairs.

\item $\hat{\kappa}M_b = \hat{\kappa}\sum_{k}{kn_k}$ is the total membrane
curvature energy in the budding regions. For a spherical geometry, the bending
energy per unit area across a NP of radius $R$ is
$\hat{\kappa} = (\kappa A/2k_BT)(2/R - c_0)^2$, 
where $\kappa$ denotes the bending modulus. Note that the spontaneous 
curvature ($c_0$) of cell membranes is nonzero. In our analysis we consider 
a vanishing spontaneous curvature ($c_0 = 0$). Therefore, $\hat{\kappa} = 2\kappa A/k_BTR^2 = 8\pi\kappa/k_BTK$. A positive spontaneous curvature ($c_0 > 0$) 
could decrease the value of $\hat{\kappa}$ while a negative spontaneous 
curvature ($c_0 > 0$) could increase the value of $\hat{\kappa}$, for a given
value of $\kappa$. We
shall discuss the effect of spontaneous curvature on cellular uptake later.
\begin{figure*}[t]
\begin{center}
\includegraphics[width=16cm]{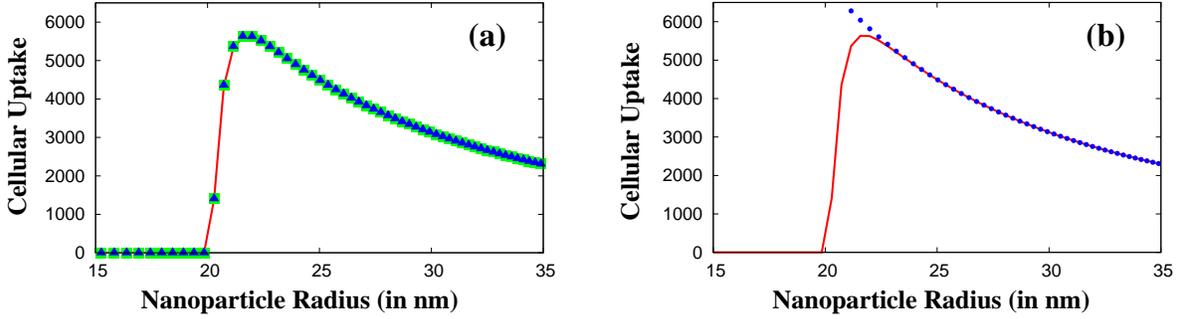}
\end{center}
\caption{Size dependent cellular uptake of non-interacting NPs.
(a) The red line is for the two-state model with $\Lambda_k = 0 = \Gamma_k$,
the green ($\square$) and blue ($\triangle$) points
are for the $K$-state system with  $\Gamma_k = 0$ and $\Gamma_k \ne 0$
respectively. (b) The red line is for the two-state model while the blue
($\bullet$) points indicate the large $R$ behaviour.}
\label{uptake-simple}
\end{figure*}

\item $\sum_{k} n_k\Gamma_k$: Total work of pulling excess membrane towards the
wrapping sites against lateral tension $\sigma$. For a single NP wrapping,
the excess area pulled towards the wrapping site is
$4\pi R^2k^2/K^2 = k^2A/K$ \cite{deserno-pre}. Thus the excess energy is
$\Gamma_k = \sigma \times {\rm excess \; area}= k^2\sigma A/k_BTK$.

\item $\sum_{k} n_k\Lambda_k$ : Total line energy of the rim, where 
$\Lambda(k)$ denotes the line energy of a {$k$}-bud. 
Assuming a spherical shape of the membrane at the rim of a partially
 wrapped NP, $\Lambda_k$ is modeled as being proportional to the length,
${\cal{L}}_k$, of its rim, with a constant line energy per unit length
$\gamma$ \cite{tzlil}. Therefore,
\begin{equation}
\Lambda_k = \gamma{\cal{L}}_k = \gamma 2\pi R\sqrt{4\frac{k}{K}\left(1 -
\frac{k}{K}\right)}.
\label{line}
\end{equation}
Note that ${\cal L}_k$ vanishes for $k = 0$ and $k = K$ and is maximum
($2\pi R$) for a half-wrapped NP ($k = K/2$). However, the local wrapping
behavior of the membrane to a NP is different from the assumption made above
\cite{deserno-bickel,deserno-pre}.
We need to consider an additional bending energy for the unadsorbed membrane
detaching from the NP at the rim. Although the $k-$dependence of this energy
is different from the simple form assumed in Eq. (\ref{line}), the general
features of large energies for half-wrapped state ($k = K/2$) and very small
energies for unwrapped and completely wrapped states are the same.

\item $w\displaystyle\sum_{k,k^{\prime}}k{k^{\prime}}n_kn_{k^{\prime}} = w \left[\sum_k kn_k \right] \left[\sum_{k^{\prime}} {k^{\prime}}n_{k^{\prime}}\right]=w M_b^2$  is the interaction
energy between the partially wrapped NPs, $w$ denoting the strength of the
interaction (or the second virial coefficient). We assume the interaction 
to depend on the degree of deformation and curvature of nearby curved 
membrane patches and
therefore to the degree of wrapping of the cell membrane to individual NPs.
Thus the total interaction energy when summed over is proportional to the 
total curved area.  
Interaction between membrane inclusions or adsorbates could arise due to a variety of different mechanisms \cite{goulian} including membrane fluctuations \cite{rg,gov}.
\end{itemize}

\noindent
In the final stages of endocytosis the membrane wrapped NP pinches off which
results in a topology change. This severing mechanism of the wrapped NP from
the membrane is brought about by proteins such as dynamin and
C-terminal binding protein 3/brefeldin A-ribosylated substrate (CtBP3/BARS).
According to Gauss-Bonnet theorem \cite{FrancoisDavid}, this leads
to an increase of $4\pi\bar{\kappa}$ in Gaussian bending
energy, with $\bar{\kappa}$ representing the Gaussian bending rigidity
of the cell membrane. In our model we are interested in events prior
to the final scission process and will therefore safely ignore this term.
In fact, all partially wrapped NPs for which $k \ge 0.9K$
will be assumed to be endocytosed.

To find the equilibrium state of the system we minimize the free energy
with respect to $L_b$ and ${n_k}$. From $\partial {\cal{F}}/\partial L_b = 0$,
we have
\begin{equation}
\frac{\phi_p}{1 - \phi_p} = \frac{\phi_b}{1 - \phi_b} \; e^{-\epsilon}.
\label{minimizelb}
\end{equation}
Minimizing $\cal{F}$ subject to the constraint of Eq. (\ref{eq:N}), we get the
normalized wrapping size distribution as
\begin{equation}
p_k = \frac{n_k}{N} = \frac{e^{-\beta_k}\alpha^k}
{\sum_{k=0}^K e^{-\beta_k}\alpha^k},
\label{dist}
\end{equation}
where we have defined
\begin{eqnarray}
\alpha = (\phi_p/\phi_b)\; e^{\epsilon - \hat{\kappa}},
\end{eqnarray}
and
\begin{eqnarray}
\nonumber
\beta_k &=& \Lambda_k + \Gamma_k + 2wk\sum_{k^{\prime}}{k^{\prime} 
n_{k^{\prime}}} \\
&=& \Lambda_k +  \Gamma_k 
+ 2 w c M k \sum_{k^{\prime}}{k^{\prime} p_{k^{\prime}}}.
\end{eqnarray}
The conservation condition for the receptors gives
\begin{equation}
\phi_p(1 - c \sum_{k} kp_k) + \phi_b c \sum_{k} kp_k = \phi_0.
\label{conserve}
\end{equation}
The densities of the receptors $\phi_p$ and $\phi_b$ can be obtained
by numerically solving Equations (\ref{minimizelb})-(\ref{conserve}).
Substituting $\phi_p$ and $\phi_b$ back into Eq. (\ref{dist}) yields the
wrapping size distribution, and hence the number of fully internalized NPs
\begin{equation}
n_K = cMp_K.
\label{num-up}
\end{equation}
Therefore, $n_K$ gives the cellular uptake of nanoparticles.
We first study the effect of particle size on the cellular uptake 
of non-interacting NPs and then see the effect of interactions on 
the distribution.

The choice of the physical constants is mostly guided by experimental data
although there are some free parameters as well. The bending modulus ($\kappa$)
of biomembranes is typically on the order of $20 \; k_{\rm B}T$
\cite{helfrich,sackmann}. The receptor-ligand binding energy, $\epsilon$ is
assumed to be comparable to antibody-antigen interaction and is estimated
to be on the order of $15-25 \; k_{\rm B}T$ \cite{nelson-cox,bell}. The length scale
is set by the length of the receptor ($\sqrt{A}$) which is typically on
the order of $15$ nm. Therefore, $A \sim 225 \; {\rm nm}^2$. Experimental
information suggests that the number of receptors varies from $50-500$
per $\mu {\rm m}^2$ \cite{tzlil,suresh,briggs,quinn}. This implies that $\phi_0$
could vary from $0.01$ to $0.1$. The concentration of NPs, $c$,
can vary between $0.001$ and $0.005$. 
The diameter of the cell being $\approx 15\mu m$, the surface
area of the cell is $\approx 707 \mu m^2$. Therefore, $M = 3.14 \times 10^6$.
In our numerical analysis, we choose
$\kappa = 20\; k_{\rm B}T, \epsilon = 25\; k_{\rm B}T, c = 0.003, \phi_0 = 0.05
$, and $M = 3.14 \times 10^6$ \cite{tzlil,suresh}. In what follows, we choose
$\sigma = 0$ and consider the effect of $\sigma$ on uptake in a later section.
Both $\gamma$ and $w$ are free variables and we choose $\gamma = 1.0$ 
(in units of $k_B T$ per unit length, $\sqrt{A}$).
$w$ is varied from zero (non-interacting) to positive (repulsion) and
negative (attraction) values.

\section{Two-State Model}
\noindent
To analyze the size-dependent uptake of NPs we make a major simplification
in the model. We assume that the NPs upon arrival to the cell surface are either
endocytosed completely or remain free without there being any intermediate
wrapped state. Then the model essentially reduces to a two-state model with the
two states being $k = 0$ and $k = K$. Our goal is to come up with the minimal
model to understand the experimental uptake behavior and to find the only
relevant energetic contributions. Note that the line energy term ($\Lambda_k$)
automatically vanishes with this simplifying assumption. We now study the uptake 
behavior (i) in the absence of interactions and (ii) when the NPs interact.
\begin{figure*}[!t]
\begin{center}
\includegraphics[width=16cm]{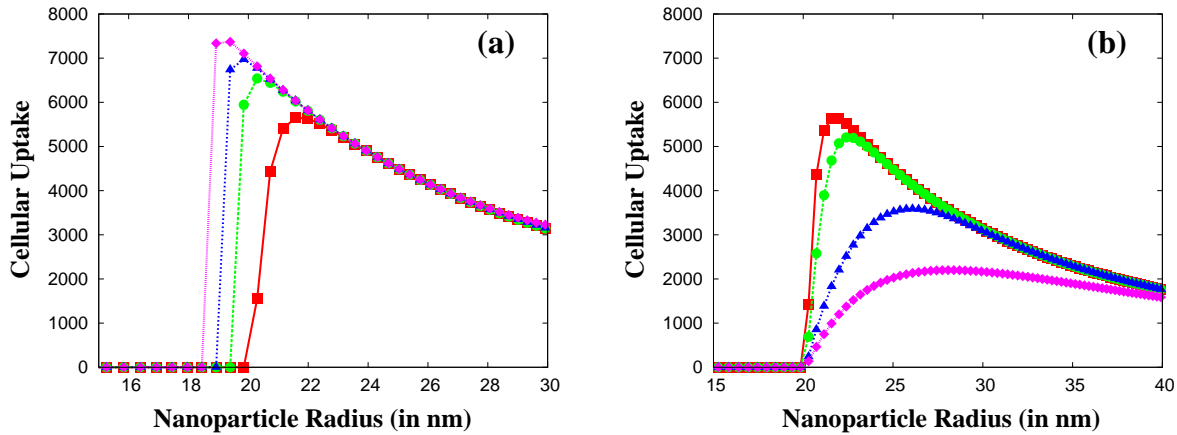}
\end{center}
\caption{Size dependent cellular uptake of interacting NPs. (a) The curves
are for different values of $w = 0.0$ (red, $\square$), $w = -0.00002$
(green, $\circ$), $w = -0.00003$ (blue, $\triangle$), and $w = -0.00004$
(pink, $\diamond$). $R_{\textrm{min}}$ decreases with increasing
strength while the optimal uptake increases. (b) The curves
are for different values of $w = 0.0$ (red, $\square$), $w = 0.00001$
(green, $\circ$), $w = 0.00005$ (blue, $\triangle$), and $w = 0.0001$
(pink, $\diamond$). Optimal uptake decreases with increasing repulsion.
}
\label{uptake-int}
\end{figure*}

\subsection{(i) Non-interacting case ($w = 0$)}
\noindent
With the two-state model, in the absence of interactions and with $\Lambda_k = 0 = \Gamma_k$,
we observe [Fig. \ref{uptake-simple}(a)] that
below a critical radius, $R_{\textrm{min}}$, there is hardly any uptake.
Above $R_{\textrm{min}}$, the uptake increases sharply to reach a maximum
and then decays as a power law with increasing radius. Thus, the two-state
model correctly reproduces the optimal uptake behavior at intermediate
radii seen in experiments. We compare our results for the two-state model with
the full K-state model both in the presence and absence of the $\Gamma_k$ term.
We find that for $\gamma = 1$ and $\sigma = 0.001$ (in units of $k_BT$ per 
unit area) the surface tension term does not affect
the size-dependent distribution significantly. 
As we shall show later, increasing $\sigma$ decreases uptake significantly 
although the uptake characteristics remain unchanged.
Thus we conclude 
that the only relevant energy terms in understanding the size-dependent 
endocytosis of NPs are the energy released on ligand-receptor binding 
and the energy cost in bending the membrane. We verify this further by 
analyzing the behavior at large and small radii.
\vskip 0.5cm 
\noindent
{\em{\bf{Behavior at large $R$.}}}
For the two state model, Eq. (\ref{conserve}) (the conservation condition) reduces
to giving the completely wrapped particle size distribution as
\begin{eqnarray}
p_K = \left[ \frac{\phi_0 - \phi_p}{\phi_b - \phi_p}\right]\frac{1}{cK}
\end{eqnarray}
In the large $R$ limit (or large $K$ limit), the
receptor density in the fully enveloped NPs is almost saturated,
$\phi_b \approx 1$, whereas the free receptor density is negligible
$\phi_p \approx 0$. Then
\begin{eqnarray}
p_K \approx \frac{\phi_0}{cK} = \frac{\phi_0A}{4\pi cR^2}.
\end{eqnarray}
Therefore, cellular uptake for larger NPs is inversely proportional to the
square of the size of the NPs and reproduces the numerically predicted
behavior at large radius exactly [Fig. \ref{uptake-simple}(b)].

\vskip 0.5cm 
\noindent
{\em{\bf{Behavior at small $R$.}}}
To understand the low cellular uptake at smaller radii we follow Tzlil et al.
to make the {\em macroscopic (bud) phase approximation},  i.e. assume that
instead of the curved regions being made up of several NPs wrapped to
different extents, there is a single NP with wrapped area $M_b$ that coexists
with the planar membrane phase.
This approximation causes the configurational entropy of the NPs in the free energy
expression to drop off. Also $\Lambda_k = 0$ for all $k$. Minimizing the
resulting free energy with respect to $L_b$ gives Eq. (\ref{minimizelb})
and minimizing it with respect to $M_b$ yields
\begin{eqnarray}
\frac{1}{1 - \phi_p} = \frac{1}{1 - \phi_b}\;e^{-\hat\kappa}.
\label{minimizemb}
\end{eqnarray}
Solving Eqs. (\ref{minimizelb}) and (\ref{minimizemb}) we can determine
the receptor densities in the two coexisting phases as \cite{tzlil}
\begin{eqnarray}
\phi_b = \frac{1-e^{-\hat\kappa}}{1-e^{-\epsilon}}\;\;\; \mbox{and}\;\;\;
\phi_p = \frac{e^{\hat\kappa}-1}{e^{\epsilon}-1}.
\end{eqnarray}
Therefore, we can have coexistence between the planar and wrapped phases only if
$\epsilon \ge \hat\kappa \ge 0$. Thus, for a single wrapped NP,
$\epsilon = \hat\kappa$ is the critical value below which we cannot have
wrapping. Substituting for $\hat\kappa$, we get the critical radius for the
onset of wrapping as \cite{lipo},
\begin{eqnarray}
R_{\textrm{min}} = \sqrt{2\kappa A/\epsilon}
\end{eqnarray}
For the values of $\kappa$ and $\epsilon$ used in the numerical estimates
we get $R_{\rm min} \approx 19$ nm.

\subsection{(ii) Interacting case ($w \ne 0$)}
\noindent
As observed above, the uptake of NPs in the absence
of interactions is highly asymmetric and also predicts a lower radius cut-off.
Experimentally, the distribution has found to be rather symmetric both for 
Au nanoparticles \cite{chitrani1,chitrani2} and DNA wrapped
single-walled carbon nanotubes (DNA-SWNT) \cite{strano1}.
Moreover, there is significant internalization
of particles below the minimum radius predicted by the model. Experiments
using DNA-SWNT show an increase in near-infrared fluorescence from SWNT
concentrated at the external cell membrane during the early stages of
endocytosis mechanism \cite{strano1}, indicating possible clustering of
nanotubes on the cell surface prior to uptake. To incorporate clustering
in our model, we include an effective interaction in our model. The idea
is that interactions could lead to clustering which could drive wrapping of
NPs of smaller sizes. In our model, the interaction between NPs is
controlled by the interaction parameter, $w$. Negative $w$ implies attraction
while positive $w$ implies repulsion. We do our analysis for the two-state
model.

{\bf{Results for attraction ($w < 0$).}} Figure \ref{uptake-int}(a) shows
the results for size-dependent uptake of NPs in the presence of attractive
interaction. Similar to the behavior in the absence of interactions, we find
that below $R_{\rm min}$, there is hardly any endocytosis. Above $R_{\rm min}$, the
uptake increases rapidly, and subsequently reaches a maximum and then decays slowly.
We note, however, that the uptake mechanism is strongly dependent on the
value of $w$. With the increase in the strength of the attractive interaction,
(increasing $|w|$), the minimum radius for uptake, $R_{\rm min}$, decreases 
substantially. Also the maximum uptake increases with increasing $|w|$ 
indicating that uptake becomes more favorable in such circumstances. 
To explain this behavior we need to look carefully at the energetics. 
Without interactions, the low radius cut-off is determined by the competition 
between the energy released on ligand-receptor binding and the energy 
cost in bending the membrane. Attractive interactions cluster NPs and 
lower the free energy locally so that the system can afford to pay the cost 
of membrane bending energy, resulting in endocytosis for smaller particle 
sizes that would otherwise have been prohibited.

{\bf{Results for repulsion ($w > 0$).}} Figure \ref{uptake-int}(b) shows
the results for size-dependent uptake of NPs in the presence of repulsive
interaction. We find that the behavior is distinctly different from that
with attractive interactions. Although the uptake mechanism strongly
depends on the interaction strength $w$, the characteristics differ
significantly. Interestingly, the lower radius cut-off $R_{\rm min}$ does not
shift with increasing $w$ and is the same for all $w$ values. The maximum
uptake however decreases with increasing $w$. The uptake behavior at
large $R$ is also affected. Although the uptake decreases at large $R$,
the decay is much slower as $w$ increases. This behavior can again be
explained by looking at the energetics. Repulsion between NPs (positive $w$),
pushes them apart and the energetics at lower radius are governed by the same
single NP wrapping energetics as in the absence of interactions. Therefore,
$R_{\rm min}$ does not change when we increase $w$. However, the uptake for
$R > R_{\rm min}$ is affected since positive $w$ increases the {\em global} free
energy cost, thus decreasing the optimal uptake and affecting the
behavior at large radius making the uptake more symmetric with particle size.

Therefore, we find that interactions affect cellular uptake of NPs significantly.
We studied the two cases of attractive and repulsive interactions separately.
However, in the physical system, we expect both of these interactions to be
present simultaneously with attractive interaction affecting the uptake of
smaller particles and repulsion dominating for the bigger ones.
Such a picture could provide a qualitative understanding of the experimental 
observation of size-dependent cellular uptake of NPs.

\subsection{Effect of membrane tension and spontaneous curvature}
\begin{figure}[h]
\begin{center}
\includegraphics[width=9cm]{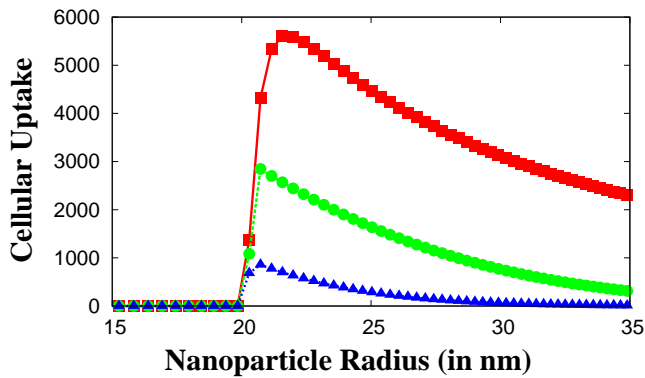}
\end{center}
\caption{Size dependent cellular uptake of non-interacting NPs for
different values of $\sigma = 0.001$ (red, $\square$), $\sigma = 0.02$
(green, $\circ$) and $\sigma = 0.1$ (blue, $\triangle$).
}
\label{st}
\end{figure}

\noindent
The effect of membrane tension is to lower the optimal uptake of NPs
\cite{yuan1,yuan2}. In
Fig. \ref{st} we show the variation of uptake with system size for different
values of the surface tension $\sigma$ (in units of $k_B T$ per unit area) in 
the absence of interactions.
The nature of cellular uptake with the particle size remains the same for
different values of $\sigma$ although the amount of uptake decreases.

In all our analysis we have ignored the effect of spontaneous curvature 
($c_0$). As we have mentioned earlier, a non-zero $c_0$ could lower or raise
the value of $\hat{\kappa}$ which should in turn lower or raise the free energy 
barrier for the uptake of a nanoparticle. Hence the optimal uptake could
indeed be lower or higher depending on the value of $c_0$. However, 
the characteristics of uptake behavior is not altered. 

\section{Summary and Conclusion}
\noindent
In this paper, we have studied the effect of interactions between surface-bound 
NPs on their subsequent endocytosis in the context of
a statistical thermodynamic model. One of our first results is to clearly
show the relevant energy terms required to understand the uptake mechanism
qualitatively. To do so, we simplified our model to a two-state model, which
captures the essential features of the uptake behavior. We showed that
apart from the important entropic contributions coming from the distribution
of the receptors and the NPs on the cell surface, the energy terms that
dictate the uptake characteristics are the energy released via ligand-receptor
binding and the energy cost in bending the membrane. Although we show that
the line energy and energy in pulling excess membrane area are not significant 
when the aim is to understand the specific uptake behavior, we do not rule out 
the importance of these terms in the endocytic process. These energy terms, 
as well as the Gaussian bending energy during the pinch-off, are relevant 
for the studies of membrane invaginations and wrapping. However, they may be
dispensable when addressing the question of the number of NPs endocytosed at a
given time.

Beyond the two-state model, our results show that interactions between NPs 
could have a drastic effect on the uptake process. 
Attractive interactions lead to clustering of NPs, which effectively
lowers the free energy threshold for wrapping and therefore shifts 
the lower cut-off radius. This is not possible in the absence of interactions 
unless, of course, we changed the relative values of $\kappa$ and $\epsilon$. 
For fixed $\kappa$ and $\epsilon$, we see that repulsive interactions also 
have a significant impact, in that they cause the cellular uptake to be reduced 
and modify uptake characteristics towards being more symmetric. 

In our model, interactions are assumed to be proportional to the wrapped area 
and the strength of the interaction, $w$, is varied freely. Therefore, $w$ 
could be thought of as an effective parameter which models the membrane-mediated 
forces \cite{goulian}. M\"{u}ller et al. \cite{muller1,muller2} study the 
curvature-mediated interaction between particles on the cell surface. 
From a purely geometric analysis, they showed that the net force between 
the particles is due to a competition between the force associated with 
the curvature along the direction joining the particles (which leads to repulsion) 
and the force associated with the curvatures perpendicular to it (which leads to attraction). 
It will be interesting to use this kind of model for membrane-mediated interaction
in a more detailed study of receptor-mediated endocytosis of interacting NPs.

\section{Acknowledgements}
The authors acknowledge fruitful discussions with the members of the 
ViNCeNS (virus like nanoparticles for targeting the central nervous system) 
project. This work was supported by grant EP/G062137/1 from the 
Engineering and Physical Sciences Research Council, United Kingdom. 


\end{document}